\documentclass[sigconf, nonacm]{acmart}

\newcommand\vldbdoi{XX.XX/XXX.XX}

\usepackage{todonotes}
\usepackage{amsthm}
\usepackage{amsmath}
\usepackage{multirow}
\usepackage{graphicx}
\usepackage{enumitem}
\usepackage{bm}
\usepackage{url}
\usepackage{hyperref}
\usepackage{subfiles}

\usepackage{caption}
\usepackage{subcaption}
\usepackage{comment}
\usepackage{float}
\usepackage[ruled,lined,linesnumbered]{algorithm2e}
\usepackage{booktabs}
\usepackage{xcolor}

\def\BibTeX{{\rm B\kern-.05em{\sc i\kern-.025em b}\kern-.08em
    T\kern-.1667em\lower.7ex\hbox{E}\kern-.125emX}}
\DeclareMathOperator*{\argmax}{arg\,max}

\newcommand{\squishlist}{ 
   \begin{list}{$\bullet$}
    { \setlength{\itemsep}{0pt}      \setlength{\parsep}{3pt} 
      \setlength{\topsep}{3pt}       \setlength{\partopsep}{0pt}
      \setlength{\leftmargin}{1.5em} \setlength{\labelwidth}{1em}
      \setlength{\labelsep}{0.5em} } }
\newcommand{\squishend}{
    \end{list}  }

\begin{document}

\title{ScienceBenchmark: A Complex Real-World Benchmark for Evaluating
Natural Language to SQL Systems}

\author{Yi Zhang}
\affiliation{%
  \institution{Zurich University of Applied Sciences}
  \streetaddress{}
  \city{}
  \state{Switzerland}
  \postcode{}
}
\email{}

\author{Jan Deriu}
\affiliation{%
  \institution{Zurich University of Applied Sciences}
  \streetaddress{}
  \city{}
  \state{Switzerland}
  \postcode{}
}
\email{}

\author{George Katsogiannis-Meimarakis}
\affiliation{%
  \institution{Athena Research Center}
  \streetaddress{}
  \city{}
  \state{Greece}
  \postcode{}
}
\email{}

\author{Catherine Kosten}
\affiliation{%
  \institution{Zurich University of Applied Sciences}
  \streetaddress{}
  \city{}
  \state{Switzerland}
  \postcode{}
}
\email{}

\author{Georgia Koutrika}
\affiliation{%
  \institution{Athena Research Center}
  \streetaddress{}
  \city{}
  \state{Greece}
  \postcode{}
}
\email{}

\author{Kurt Stockinger}
\affiliation{%
  \institution{Zurich University of Applied Sciences}
  \streetaddress{}
  \city{}
  \state{Switzerland}
  \postcode{}
}
\email{}

\begin{abstract}
Natural Language to SQL systems (NL-to-SQL) have recently shown {\color{black}improved accuracy (exceeding $80\%$) for natural language to SQL query translation}  due to the emergence of transformer-based language models, and the popularity of the Spider benchmark. However, Spider mainly contains simple databases with few tables, columns, and entries, which do not reflect a realistic setting. Moreover, complex real-world databases with domain-specific content have little to no training data available in the form of NL/SQL-pairs leading to poor performance of existing NL-to-SQL systems. 

In this paper, we introduce \emph{ScienceBenchmark}, a new complex NL-to-SQL benchmark for three real-world, highly domain-specific databases. For this new benchmark, SQL experts and domain experts created high-quality NL/SQL-pairs for each domain. To garner more data, we extended the small amount of human-generated data with synthetic data generated using GPT-3. We show that our benchmark is highly challenging, as the top performing systems on Spider achieve a very low performance on our benchmark. Thus, the challenge is many-fold: creating NL-to-SQL systems for highly complex domains with a small amount of hand-made training data augmented with synthetic data. To our knowledge, \emph{ScienceBenchmark} is the first NL-to-SQL benchmark designed with complex real-world scientific databases, containing challenging training and test data carefully validated by domain experts.  
\end{abstract}

\maketitle



\section{Introduction}



Enabling users to query structured  data using natural language is considered the key to data democratization. Natural Language Interfaces for Databases (or NL-to-SQL systems) emerged in the 1970s \cite{DBLP:journals/nle/AndroutsopoulosRT95, affolter2019survey}. Early systems relied on the database schema to build a SQL query from a natural language (NL) query (e.g., SODA \cite{blunschi2012soda}, Precis \cite{simitsis2008precis}) or focused on understanding the structure of the natural language query to map it to SQL (e.g., ATHENA~\cite{saha2016athena}, NaLIR~\cite{li2014nalir}).
As early as 1995, the lack of benchmarks was apparent: ``No standard benchmarks have yet been developed [...], any appraisal of the current state of the field must be impressionistic'' \cite{DBLP:journals/nle/AndroutsopoulosRT95}. This situation changed recently, when the first large-scale benchmarks, WikiSQL \cite{zhong2017seq2sql} and Spider \cite{yu-etal-2018-spider}, emerged. These allowed for training and evaluating \emph{neural machine translation (NMT) approaches}   (e.g., \cite{zhong2017seq2sql, xu2017sqlnet, wang:ratsql}). These approaches formulate the NL-to-SQL problem as a language translation problem, and train neural networks with large amounts of NL/SQL-pairs.

While the first deep learning approaches \cite{zhong2017seq2sql, xu2017sqlnet} only worked for single tables and failed to generate complex SQL queries spanning multiple tables (e.g., including nested queries and complex clauses), recent systems \cite{deriu-etal-2020-methodology,wang:ratsql,brunner2021valuenet, scholak2021picard} work on complete databases and achieve high performance scores on the Spider benchmark \cite{yu-etal-2018-spider}.
The top NL-to-SQL systems reach accuracies up to 85\% on Spider. However, the majority of databases present in Spider were created specifically for this benchmark and are not representative of the difficulties that arise when creating an NL interface for a real-world database.
Among the current best-performing, open-source, systems on Spider,   T5-Large \cite{raffel2019t5} (with Picard \cite{scholak2021picard} for constrained decoding), SmBoP \cite{rubin-berant-2021-smbop-semi} and RESDSQL with NatSQL \cite{DBLP:conf/aaai/Li00023}, are already achieving over $70\%$ performance.
Applying a system trained on the Spider dataset to a new domain such as astrophysics or cancer research, yields poor results, making the adoption of such systems to real-life applications extremely far-fetched.

\begin{figure*}
    \centering
  \includegraphics[width=0.99\textwidth]{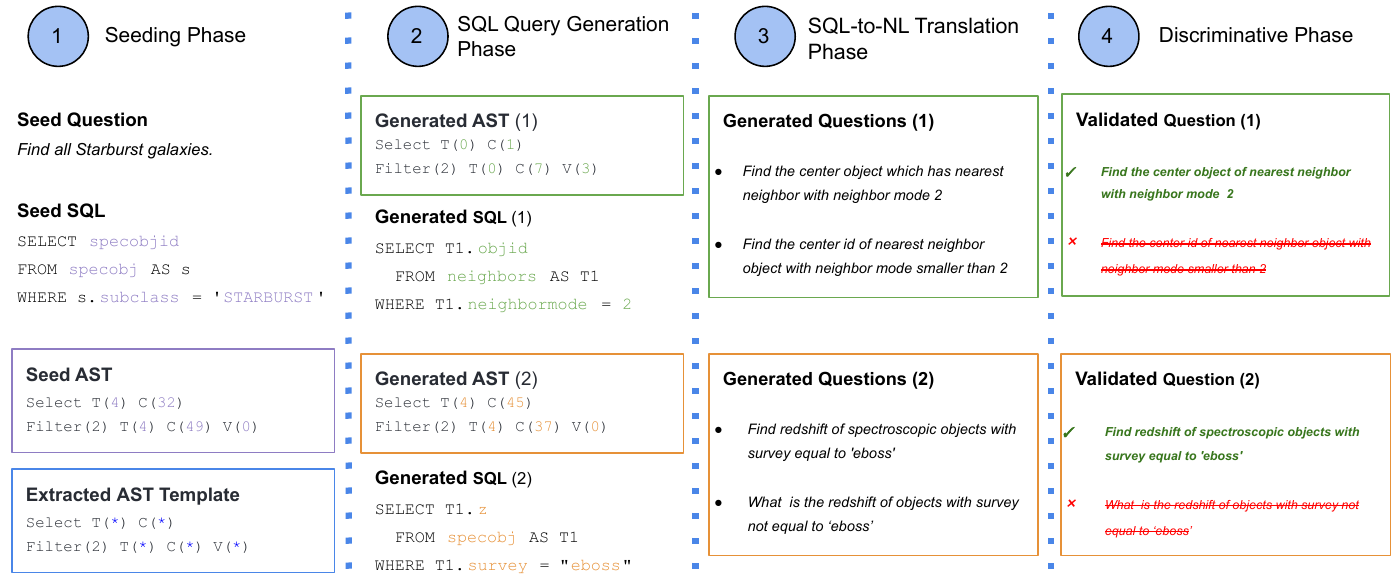}
  \caption{End-to-end architecture for automatic training data generation consisting of four different phases, namely (1) Seeding Phase, (2) SQL Generation Phase, (3) SQL-to-NL Translation Phase and (4) Discriminative Phase. The approach is used to produce our novel benchmark dataset called ScienceBenchmark. The details of these phases are described in Section \ref{sec:automatic-generation}.}
  \label{fig:data-flow-Working-pipeline1}
\end{figure*}

The problem is that, while WikiSQL and Spider provide a common training and evaluation tool that has been a game changer for the development of NL-to-SQL systems, they have serious limitations.
Each  WikiSQL question is directed to a single table and not to a relational database. The low complexity of the WikiSQL queries makes its value for real-life applications practically limited.
Spider contains 200 relational databases from 138 different domains along with over 10,000 natural language questions and over 5,000 complex SQL queries, hence it targets the query complexity problem. At the same time, it also attempts to solve the transfer learning problem  by providing a dev set with novel databases. 

However, the majority of databases in the Spider benchmark \emph{are rather simple}, i.e.,  of low complexity, and containing only what is considered general knowledge.
Thus, we need specialized, domain-specific benchmarks for training and evaluating NL-to-SQL systems in scientific domains. Manually crafting such a benchmark is prohibitive due to the volume of data needed and the expertise required in scientific domains. \emph{Data augmentation}, i.e. automatic benchmark generation is the only feasible solution.

\textbf{Our Approach}.
In this paper, we introduce \emph{ScienceBenchmark}, a complex real-world benchmark for evaluating NL-to-SQL systems. It is the first of its kind to be developed collaboratively with SQL experts and researchers from the fields of research policy making, astrophysics, and cancer research. We \textit{combine} the knowledge derived from \textit{manual training data collection with an automatic training data generation system} that enables NL-to-SQL systems that require large amounts of data for training to be bootstrapped on complex, scientific databases where training data would otherwise be scarce or unavailable. 

The main architecture of the data augmentation system is described in Figure \ref{fig:data-flow-Working-pipeline1}. 
{\color{black} The system is fed a \textit{small set} of manually generated NL/SQL-pairs to provide accurate and highly semantically relevant information about a novel domain. 
\textit{SQL templates are extracted} from these seed queries, and are subsequently used to generate SQL queries over a specific database taking into account the database schema and contents. }
These SQL queries are then \emph{back-translated to natural language} using the large language model {\color{black}(LLM)}, GPT-3~\cite{brown2020gpt3}. For the SQL-to-NL component used in this work, we experimented with state-of-the-art transformer-based pre-trained models that have shown their NL generation capabilities by achieving state-of-the-art scores in multiple related tasks.
Our evaluation showed that the GPT-3 \cite{brown2020gpt3} model achieves the best performance and is able to generalize to new domains with very few samples {\color{black}  (in our case, these are the seed NL/SQL-pairs),} which is why it was integrated in our architecture.
The resulting natural language questions are then filtered using a  \emph{critic model} to select the most relevant NL question for the corresponding SQL query. 
{\color{black}The resulting augmented dataset of \emph{ NL/SQL-pairs} can be fed into any NL-to-SQL system for training. Our approach is generic and can boost the accuracy of any NL-to-SQL system as demonstrated in our experiments in Table \ref{tab:augment2}.}


\textbf{Contributions}. The major contributions of our work are the following:
\squishlist
    \item We contribute \emph{ScienceBenchmark - a new benchmark for evaluating NL-to-SQL systems against complex, scientific databases}. \emph{ScienceBenchmark} contains more than 6,000 NL/SQL-pairs to help researchers address the complex challenges of real-world databases, overlooked by popular benchmarks.
    \item We have built \emph{ScienceBenchmark} using a \textit{novel approach for automatically generating training data for databases} where little to no training data exists. Unlike previous work that focused on rather simple databases, we \textit{concentrate on complex, real-world scientific databases}, an area where popular NL-to-SQL systems typically falter due to their lack of domain knowledge and training data.
    \item We {\color{black}evaluated three state-of-the-art NL-to-SQL systems as well as two LLMs on our benchmark}. Although these systems achieve top scores on the Spider leaderboard (above {\color{black}82\%} accuracy), none achieves a satisfactory score on our benchmark (only {\color{black} in the range of 25-60\%} accuracy depending on the domain), showcasing the difficulty of \emph{ScienceBenchmark}. 
\squishend

\section{The Need for a New Benchmark - Motivating Example: Astrophysics}
\label{sec:running_example}

A NL-to-SQL system needs to address many challenges \cite{androutsopoulos1995nlidb,affolter2019survey,iacob2020neuralsurvey,katsogiannis2023survey}. On the one hand, a natural language question may be vague, contain references that are even hard for a human to understand, and use a different vocabulary from the one used in the DB. On the other hand,  the respective SQL query needs to adhere to a strict syntax and to the underlying DB schema {\color{black} in order to be syntactically and semantically correct. 
When applying an NL-to-SQL system to a real-world scientific database, additional challenges arise that stem from the nature and the domain of these databases.}
While the Spider benchmark is the first large-scale dataset with complex SQL queries, its databases cannot be considered complex. Their subject-matter is very generalized, covering topics such as  pets and entertainment (concerts, orchestras, musicals etc.). The majority of these databases were created by students specifically for Spider and are not  representative of real-world databases.

In what follows, we motivate the need for a novel design and training of NL-to-SQL systems for complex, scientific databases. We will focus our discussion on astrophysics, a very data-intensive and highly complex scientific discipline with a long tradition of using relational databases \cite{szalay2002sdss}. The challenges described here are not only relevant for astrophysics but also for other scientific disciplines, such as cancer research, also included in our experimental evaluation. 

As our running example, we use the astrophysics database called Sloan Digital Sky Survey (SDSS)\footnote{https://www.sdss.org/}. This database stores information about stars and galaxies at specific locations in the sky. Further details on the complexity of this database are specified in Section \ref{sec:datasets}.

Let us consider three different representative astrophysics queries that serve as running examples for our paper.

\squishlist
    \item \emph{Q1: Find all Starburst galaxies.}

    \item \emph{Q2: What is the object id, right ascension, declination and redshift from spectroscopically observed galaxies with redshift greater than 0.5 but less than 1?}

    \item \emph{Q3: Find the photometric objects with object ids and spectroscopic object id whose spectroscopic class is 'GALAXY', with the difference of magnitude u and magnitude r less than 2.22 and the difference of magnitude u and magnitude r greater than 1.}
\squishend

Their corresponding SQL queries are as follows:

\begin{footnotesize}
\begin{verbatim}
    Q1: (Spider hardness: Easy)
    SELECT s.specobjid
    FROM specobj AS s
    WHERE s.subclass = 'STARBURST'
    
    Q2: (Spider hardness: Medium)
    SELECT s.bestobjid, s.ra, s.dec, s.z
    FROM specobj AS s
    WHERE s.class = 'GALAXY' AND s.z > 0.5 AND s.z < 1
    
    Q3: (Spider hardness: Extra hard)
    SELECT p.objid, s.specobjid
    FROM photoobj AS p
      JOIN specobj AS s ON s.bestobjid = p.objid
    WHERE s.class = 'GALAXY'
      AND p.u - p.r < 2.22 AND p.u - p.r > 1
\end{verbatim}
\end{footnotesize}


As shown by these queries, the major challenges for NL-to-SQL systems for complex, scientific databases are as follows:

\squishlist
    \item \emph{Unseen domains}: To understand a \textit{challenging domain} like astrophysics and write meaningful queries - in natural language and SQL - extensive domain knowledge is required. Hence, neural machine translation systems pre-trained on common knowledge datasets, like Spider, typically fail in complex domains due to the large disparity in subject matter. 

   \item \emph{Complex, often cryptic, database schemas}:  Scientific databases often store large amounts of data with \emph{hundreds of attributes}. Moreover, attributes may have extremely short names, such \textsl{ra} or \textsl{z} (referring to right ascension and redshift, respectively, in astrophysics) or store numerical measurements. Hence, additional ontologies that describe the meaning or the scientific interpretation of these attributes - which can be complex astrophysics descriptions including mathematical equations and natural language texts - are required. Finally, learning the mapping of a token from a natural language question to the relevant database attribute is non-trivial. 

   \item \emph{Sophisticated queries}: Domain-specific queries may be more elaborate than the ones in Spider. For instance, astrophysics analysis requires the use of \emph{functions} and \emph{mathematical operators} between attributes, such as the difference in magnitude between ultraviolet (\textsl{u}) and infrared filters (\textsl{r}), e.g., \textsl{u - r $<$ 2.22}. 
\squishend


These observations expose the need for specialized benchmarks designed to capture the particularities and semantics of the domain at hand as well as the types of queries that need to be understood by an NL-to-SQL system. These requirements, in combination with the size that such a benchmark necessitates, prohibit its manual construction.

Since large and complex schemas are hard (even for experts) to understand and query, the question that naturally arises is \emph{how to build such benchmarks?} The answer is data augmentation, which in turn brings its own challenges: {\color{black} \emph{How to build representative SQL queries for a new database? How to come up with their NL descriptions? How to do it with minimal, if any, human involvement? }}

\section{Science Benchmark: A New Benchmark for Complex Databases}
~\label{sec:datasets}
In this section, we present \emph{ScienceBenchmark}, which is composed of three domain-specific databases, namely research policy making, astrophysics and cancer research. First, we introduce the databases, showing their complexity and the characteristics of each domain. Second, we describe the manual data collection, which includes SQL experts writing queries with the involvement of domain experts as part of multi-year research project called INODE - Intelligent Open Data Exploration \cite{amer2022inode}. Third, we describe our automatic data augmentation approach for generating synthetic training data using GPT-3~\cite{brown2020gpt3}. Finally, we show the training and evaluation datasets of our novel benchmark \emph{ScienceBenchmark}.

\subsection{Complex, Real-World Databases} \label{sec:databases}
Here, we introduce complex, real-world databases from three scientific domains, which are all of significantly greater size and complexity than the databases found in Spider~\cite{yu-etal-2018-spider}. Table~\ref{tab:database_stats} provides an overview {\color{black} of the complexity of the Spider, KaggleDBQA\cite{lee-etal-2021-kaggledbqa}, BIRD\cite{li2023llm} and ScienceBenchmark databases. The table shows that BIRD is significantly larger and more complex than KaggleDBQA (i.e. number of tables, columns and rows). We also show that BIRD has a larger number of cross-domain databases than ScienceBenchmark. However, the complexity of each individual database of ScienceBenchmark is much higher than BIRD in the number of tables and size per database. In this sense, BIRD and ScienceBenchmark are complementary.}

\begin{table}[t]
\centering
\resizebox{0.49\textwidth}{!}{
    \begin{tabular}[h]{@{} l r r r r r r r @{}}

\toprule
     \textbf{Dataset} & \textbf{DBs} & \color{black}{\textbf{Tbls. }}& \color{black}{\textbf{Cols.}} & \textbf{Rows} & \color{black}{\begin{tabular}[x]{@{}c@{}}\textbf{Avg.}\\\textbf{R/T}\end{tabular}} & \color{black}{\begin{tabular}[x]{@{}c@{}}\textbf{Size}\\\textbf{(GB)}\end{tabular}} & \color{black}{\begin{tabular}[x]{@{}c@{}}\textbf{\#NL/SQL}\\\textbf{ (man.+syn.)}\end{tabular}} \\
\midrule
     Spider & 186 & 641  & 4,268 & 1.6M & 2.5K & 0.51 & \color{black}{8,053 + 0}\\
     (Avg / DB) & & 3.5 & 23 & 8.6K & & 0.03 & \color{black}{43 + 0}\\
\midrule
     \color{black}{KaggleDBQA} & \color{black}{8} & \color{black}{17}  & \color{black}{179} & \color{black}{4.7M} & \color{black}{280K} & \color{black}{0.4} & \color{black}{272 + 0} \\
     \color{black}{(Avg / DB)} & & \color{black}{2.1} & \color{black}{22.3} & \color{black}{595K} & & \color{black}{0.05} & \color{black}{34 + 0}\\
\midrule
     \color{black}{
     BIRD} & \color{black}{81} & \color{black}{604}  & \color{black}{4,456} & \color{black}{3.7B} & \color{black}{608.8K} & \color{black}{33.4} & \color{black}{12,751 + 0} \\
     \color{black}{(Avg / DB)} & & \color{black}{7.5} & \color{black}{55} & \color{black}{4.5M} & & \color{black}{0.35} & \color{black}{135 + 0} \\
\midrule
\midrule
    \color{black}{\begin{tabular}[x]{@{}l@{}}\emph{Science-} \\\emph{Benchmark}\end{tabular}}& & & & & \\
\midrule
     CORDIS & 1 & 19 & 82 & 671K & 35K & 1 & \color{black}{200 + 1306}\\
     SDSS & 1 & 6 & 61 & 86M & 14M & 6.1 & \color{black}{200 + 2061}\\
     OncoMX & 1 & 25 & 106 & 65M & 2.6M & 12 & \color{black}{200 + 1065}\\ 
\bottomrule
\end{tabular}

}
\caption{Complexity of the Spider, {\color{black} KaggleDBQA and BIRD} databases 
compared with the databases of our new benchmark dataset \emph{ScienceBenchmark}. The table shows the number of databases (DBs), tables, columns, rows per DB, average number of rows per table and DB size. The scientific domains of the databases contained in the ScienceBenchmark are \emph{research policy making} (CORDIS), \emph{astrophysics} (SDSS) and \emph{cancer research} (OncoMX).} \label{tab:database_stats}
\vspace*{-15pt}
\end{table}

\vspace{2pt}

\noindent{\bf Research Policy Making:} The CORDIS database, i.e. \emph{Community Research and Development Information Service\footnote{https://cordis.europa.eu/}}, serves as the European Commission's primary source of results from the projects funded by the EU's framework programs for research and innovation. The database contains very detailed hierarchical information about the framework of funding calls and the network of industrial and academic institutions, all of which is coded in highly specific enigmatic EU terminology. 

An example of this is the acronym NUTS, which stands for \emph{nomenclature of territorial units for statistics}. Even the long form does not necessarily give the casual user a clear idea of what kind of information might be stored in such a column. Another challenging aspect of this database is the amount of text (e.g. descriptive project objectives averaging 1,821 characters) and the diversity of topics in the database ranging from \emph{Information and Media} to \emph{Nuclear Fission}. For \emph{ScienceBenchmark}, we use version 2022-08 of the database, as shown in Table \ref{tab:database_stats}, which comprises 19 tables and 82 columns and has an average of 35,355 rows per table. The database size is 1 GB. 

\vspace{2pt}

\noindent{\bf Astrophysics:} The SDSS (\emph{Sloan Digital Sky Survey}) -- introduced in Section \ref{sec:running_example} -- is a database containing the most detailed three-dimensional map of the universe ever made. The data collection began in 2000 and continues today. The database has 10 tables that contain disparate numbers of columns varying between 3 and 804 columns each. The tables contain various measurements and information about the type of observed object (e.g. a star or a galaxy), distances between observed objects, and various parameters (e.g. right ascension, declination, photometric system filters (u, g, r, i and z)) that have been measured in photometric or spectroscopic observations.

In order to study star-forming galaxies, the sky is measured in several color bands such as infrared or ultraviolet resulting in different spectra, i.e. alternative numerical measurements and thus various interpretations of galaxies. Unlike on Earth, the location of celestial objects is defined by their right ascension and declination. Moreover, literally, hundreds of different attributes are collected for each object, such as size, redshift, brightness, magnitudes of color bands measurements, etc. This database contains many column names and values that are labeled with abbreviations (rather than natural language) that are familiar to astrophysicists but indecipherable for non-specialists. In order to query this database with an NL-to-SQL system, it was necessary to add natural language labels for abbreviated columns (e.g. ra = right ascension). In addition to the lack of natural language labels for columns, much of the information in the database is numerical and is used in complex queries with mathematical operations. Querying numerical data in natural language is significantly more complex than querying a database comprised of mostly textual values. 
    
Due to the limitation of the input tokens of the language model used in our experiments, we use a subset of the database comprised of 5 original tables and 1 additional table for photometrically observed astronomical objects. 
As described in Table \ref{tab:database_stats}, there are 61 columns, averaging about 10 columns per table and an average of 14,462,875 rows per table.  The size of the database is 6.1 GB.  \\

\noindent{\bf Cancer research data:} OncoMX\footnote{https://www.oncomx.org/} is a database funded by the U.S. National Institute of Health (NIH) that integrates knowledge from several different sources about cancer biomarkers. The version of OncoMX used in \emph{ScienceBenchmark} contains information from cancer biomarker databases (EDRN\footnote{https://edrn.nci.nih.gov/}, FDA\footnote{https://www.fda.gov/}), gene expressions in healthy anatomical entities (Bgee\footnote{https://bgee.org/}), differential gene expressions between healthy and cancerous samples (BioXpress\footnote{https://bioxpressbiosimilars.com/}) and cancer mutations (BioMuta\footnote{https://hive.biochemistry.gwu.edu/biomuta/norecord}). 
As shown in Table \ref{tab:database_stats}, the database comprises 25 tables that have 2 to 14 columns each, for a total of 106 columns and has an average of 2,636,771 rows per table. The size of the database is 12 GB. 

The complexity in this database lies in the heavily domain-specific information it contains as well as the complex queries that researchers use when exploring this database. For example, those unfamiliar with cancer research will not know that “BRCA1” or “BRCA2” refers to \emph{BReast CAncer gene 1} and \emph{BReast CAncer gene 2}, respectively. In addition to the domain-specific information, even a seemingly simple query in natural language such as "Show biomarkers for breast cancer" requires a SQL query with a multi-relational join and several filters. 

\subsection{Manual Data Collection} \label{sec:manual}
\label{sec:manuel_data_collection}
In this section, we detail the manual data collection for all three databases. All of the data was generated and reviewed by an expert group consisting of at least one SQL expert and one domain expert in research policy-making, astrophysics, or cancer research. In total, the team consisted of about 20 domain and SQL experts of various age ranges and genders. All of the experts were members of the multi-year research project INODE \cite{amer2022inode}, including partners from academia and industry.

Before starting data collection, the domain experts such as astrophysicists and cancer researchers introduced the SQL experts to the domain-specific knowledge within the database. At the data generation stage, the domain experts  developed the \emph{natural language questions}, while the SQL experts were responsible for writing the corresponding \emph{SQL queries}. 

It is important to note that the domain experts were given the task of solving realistic science questions rather than simply generating complex questions based on the database. During the review and validation phase, domain experts used their expertise to verify the SQL queries together with the output of the SQL queries. 
For each domain, we generated a {\color{black} \textit{training set} of 100 NL/SQL pairs  as well as a \textit{test set} of 100 NL/SQL pairs}. 

In contrast, for each database in the Spider dev set, there are far fewer questions, 50 on average per database, ranging from 63 to just 4 questions per database. We have double the amount of dev set queries for our databases than in Spider.
\subsection{Automatic NL-to-SQL Training Data Generation}
\label{sec:automatic-generation}
In this section, we present our automatic training data generation approach.  {\color{black}This approach is generalizable and can be applied to any domain.} We will use our running example for astrophysics introduced in Section \ref{sec:running_example}. A concrete example of the end-to-end pipeline is depicted in Figure \ref{fig:data-flow-Working-pipeline1}. 

{\color{black}The process of automatic training data generation involves four phases: 1) the \emph{Seeding Phase}, where SQL templates are extracted from the manually written seed queries, 2) the \emph{SQL Query Generation Phase}, where the templates are filled with the database content, and schema are used to create a readable version of the query,} 3) the \emph{SQL-to-NL Translation Phase}, where GPT-3 generates a set of candidate questions, and 4) the \emph{Discriminative Phase} (candidate selection phase) that selects the top two NL questions per SQL query. 

\subsubsection{Phase 1: Seeding Phase.} The seeding phase ingests the manually created SQL queries (as discussed in Section \ref{sec:manual}) and extracts \emph{query templates}. For this, the manually created queries are transformed into an \emph{Abstract Syntax Tree} (AST) representation called \emph{SemQL} \cite{guo2019irnet}. The leaf nodes of the AST, i.e., tables (T), columns (C) and values (V) are replaced with placeholders (denoted as (*) in Figure \ref{fig:data-flow-Working-pipeline1}). The resulting AST is used as a \emph{query template}, which is filled with database content in the next phase.

\subsubsection{Phase 2: SQL Query Generation Phase.} 
When populating the query templates, the usage of randomly sampled tables and columns without any constraints might lead to meaningless or unrunnable queries. In order to ensure that the generated SQL queries are meaningful and useful to researchers in each field, we automatically create an \emph{enhanced schema}. The enhanced schema can also be refined manually by domain experts to offer more meta-information about tables and columns. The manual work, if needed, is one-shot. This enhanced schema enables the exposure of the following information to the system. 

\squishlist
\item \emph{Non-Aggregatable Columns:} These columns should not be allowed to appear in an attribute with an aggregation function like \texttt{SUM}, \texttt{AVG}, \texttt{MIN}, \texttt{MAX} because such an operation is not meaningful. An example below shows a query for getting an average of all IDs of spectroscopic objects, which is executable but not meaningful.

\begin{footnotesize}
\begin{verbatim}
SELECT AVG(s.specobjid) FROM specobj as s
\end{verbatim}
\end{footnotesize}

\item \emph{Columns with Categorical Values:} Typical examples are \emph{gender} or \emph{number of languages spoken} by a person. These columns usually have a low cardinality and are more likely applied to a \texttt{GROUP BY} clause. 

The example below stands for \emph{How many spectroscopic objects are there for each right ascension?} This question will hardly be asked by anyone knowing the basics of astrophysics. The SQL-statement may return millions of rows with useless information, because of the very high cardinality of right ascension. 

\begin{footnotesize}
\begin{verbatim}
SELECT COUNT(*), s.ra 
FROM specobj as s GROUP BY s.ra
\end{verbatim}
\end{footnotesize}

With this constraint, the sampling procedure ensures that more meaningful queries are generated. The SQL-statement below retrieves the often asked question: \emph{Find the count of spectroscopic objects grouped by their corresponding class}.

\begin{footnotesize}
\begin{verbatim}
SELECT COUNT(*), s.class 
FROM specobj as s GROUP BY s.class
\end{verbatim}
\end{footnotesize}

\item \emph{Columns relevant for Applying Math Operators:} These columns are chosen by our sampling algorithm to apply certain math operators. Identifying these columns ensures that there will be no unexpected randomness among the operands, such as \textit{T1.length - T2.area} -- which is not meaningful.

\item \emph{Semantically Meaningful Table and Column Names}. Since complex, scientific databases often contain table and column names that are not easily interpretable for humans (e.g., \emph{ra} stands for \emph{right ascension}), we introduce human-readable aliases, which spell out the abbreviated table and column names. This facilitates both the automatic SQL-to-NL generation as well as the manual creation of NL/SQL pairs since the domain experts are aided with more meaningful names. 

Let us revisit the SQL-statement query Q2 of our running example in Section \ref{sec:running_example}. It is not clear, what attribute \textit{s.z} is referring to, since there is no extra information about the column \textit{z} in table \textit{specobj}. However, given the logical alias of \textit{s.z}, we are able to see that \textit{s.z} stands for redshift, which can then be rewritten as \textit{spectroscopic\_object.redshift}. Using the same transformation on all tables and columns, we obtain the readable and semantically meaningful SQL query which facilitates the development of the corresponding NL questions. 


\squishend

In the next step of the second phase, \emph{query templates} are filled with the contents of the database (i.e., table names, column names, and values) using the enhanced schema. To increase the diversity of the queries in the synthetic dataset, we apply \emph{random sampling} to the AST representation of the template. The random sampling only \emph{changes the leaf nodes in the AST}, which represent the corresponding columns, tables, and values.
For instance, the projection column \textit{specobjid} from  table \textit{specobj} may be changed to column \textit{objid} from table \textit{neighbors}, as shown by \emph{Generated AST (1)} and \emph{Generated SQL (1)} in Figure~\ref{fig:data-flow-Working-pipeline1}. 
Another result of the random sampling is represented by \emph{Generated AST (2)} and \emph{Generated SQL (2)}. In this case, the table \textit{specobj} is still used, but the projection column \emph{z}, as well as the filter condition on the column \emph{survey}, are new. 

\vspace{2pt} 

\noindent\textbf{Algorithm for SQL Query Generation}
{Algorithm \ref{algo:GenereteSQLQueriesAlgo}
\label{random_sampling} details the step by step process of automatically generating a SQL query using the AST templates and enhanced schema. 
We explain the algorithm with the example shown in Figure \ref{fig:data-flow-Working-pipeline2}. 
In particular, we analyze the generation of the following SQL query which is also shown in Figure \ref{fig:data-flow-Working-pipeline1}:

\begin{footnotesize}
\begin{verbatim}
SELECT T1. objid FROM neighbors AS T1 
WHERE T1. neighbormode = 2 
\end{verbatim}
\end{footnotesize}

\begin{figure}[h]
  \includegraphics[width=0.7\columnwidth]{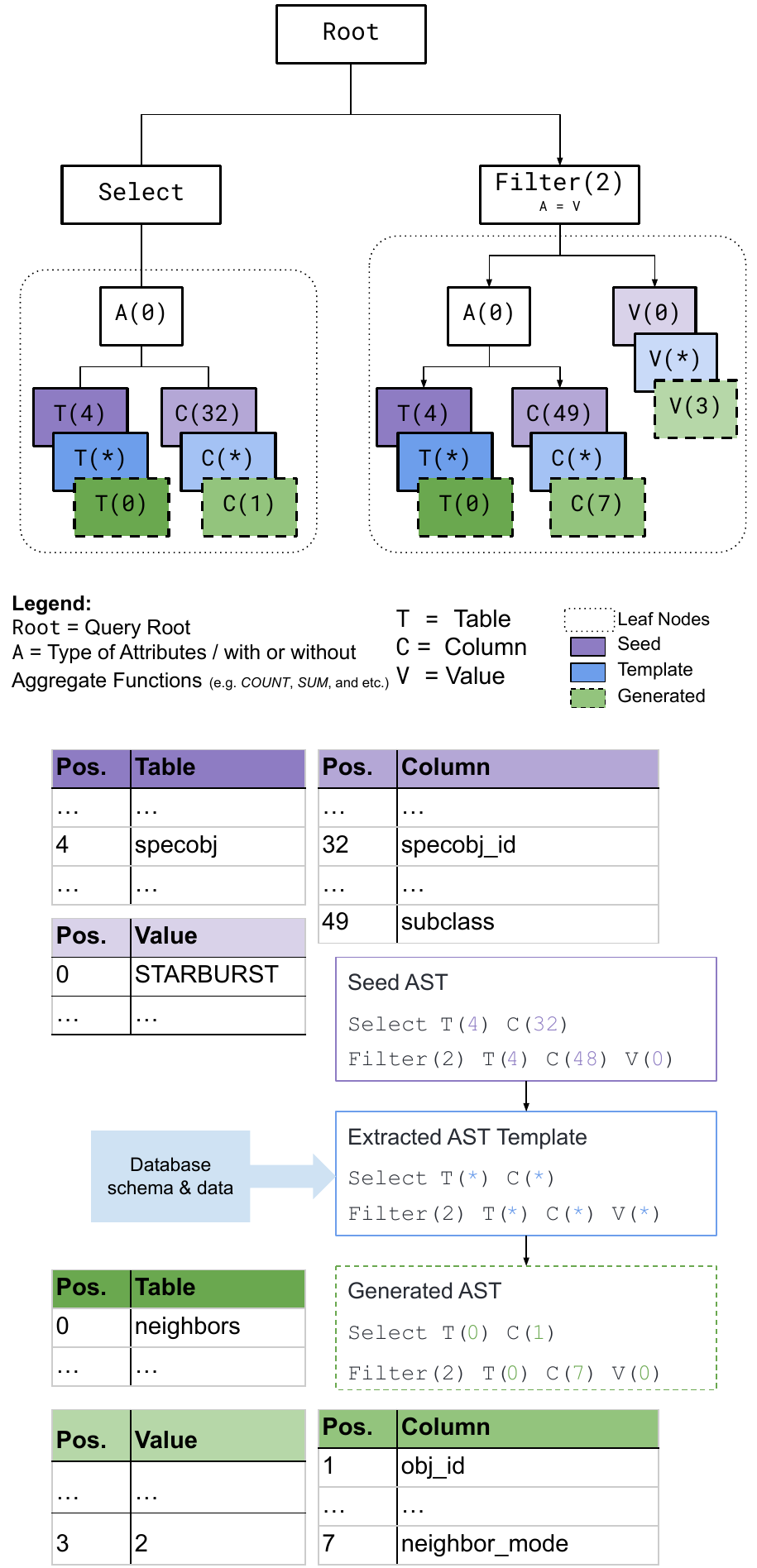}
  \caption{Example of extracting \& applying query templates for automatically generating SQL queries as shown in Algorithm \ref{algo:GenereteSQLQueriesAlgo}. The top part shows the abstract syntax tree (AST) of a specific query. The lower part shows how query templates are applied for generating the SQL queries based on database schema and data.}
  \label{fig:data-flow-Working-pipeline2}
\end{figure}

The algorithm starts with extracting leaf nodes from the AST templates and initializes a new temporary set of empty hash-maps, including Tables, Columns and Values (see lines 1 to 6).

\begin{algorithm}[h]
\small
\caption{Generating SQL queries from an AST template and the enhanced schema}
\label{algo:GenereteSQLQueriesAlgo}
\SetKwData{emptydict}{EmptyHashMap}
\SetKwData{AST}{\textit{Ast}}
\SetKwData{GenerativeSchema}{\textit{EnhancedSchema}}
\SetKwData{OutputSQL}{\textit{Sql}}
\SetKwData{LeafNodes}{\textit{LeafNodes}}
\SetKwData{Tuple}{\textit{LeafNode}}
\SetKwData{Columns}{\textit{Columns}}
\SetKwData{Tables}{\textit{Tables}}
\SetKwData{Values}{\textit{Values}}
\SetKwData{AggregatorPos}{\textit{AggregatorPosition}}
\SetKwData{ColumnPos}{\textit{ColumnPosition}}
\SetKwData{TablePos}{\textit{TablePosition}}
\SetKwData{ValuePos}{\textit{ValuePosition}}
\SetKwData{ColumnVal}{\textit{ColumnValue}}
\SetKwData{TableVal}{\textit{TableValue}}
\SetKwData{ValueVal}{\textit{ValueValue}}
\SetKwFunction{ExtractLeafNodes}{ExtractLeafNodes}
\SetKwFunction{SampleTable}{SampleTable}
\SetKwFunction{SampleColumn}{SampleColumn}
\SetKwFunction{SampleValue}{SampleValue}
\SetKwFunction{Transform}{Transform}
\SetKwInOut{Input}{Input}
\SetKwInOut{Output}{Output}

\Input{An AST Template \AST}
\Input{An enhanced schema of the target database \GenerativeSchema}
\Output{A new generated SQL \OutputSQL}

\Begin{
\LeafNodes $\leftarrow$ \ExtractLeafNodes(\AST)\;

\Tables $\leftarrow$ \emptydict\;

\Columns $\leftarrow$ \emptydict\;

\Values $\leftarrow$ \emptydict\;

\ForEach{$\Tuple \in \LeafNodes$}{
  \AggregatorPos, \TablePos, \ColumnPos, \ValuePos $\leftarrow$ \Tuple \tcp*{as quadruple} \label{algo:quardruple}
      
  \If{$\exists \TablePos : (\TablePos, \TableVal) \notin \Tables$}
  {
    \TableVal $\leftarrow$ \SampleTable(\Tables, \GenerativeSchema)\;
    
    \Tables.update(\TablePos, \TableVal)\;
  }
  \If {$\exists \ColumnPos : (\ColumnPos, \ColumnVal) \notin \Columns$}
  {
    \ColumnVal $\leftarrow$ \SampleColumn(\AggregatorPos, \TableVal, \Columns, \GenerativeSchema)\;
    
    $\Columns.update(\ColumnPos,$
    $\ColumnVal)$\;
  }
  \If{$\exists \ValuePos : (\ValuePos, \ValueVal) \notin \Values$}
  {
    \ValueVal $\leftarrow$ \SampleValue(\TableVal, \ColumnVal)\;
    
    $\Values.update(\ValuePos, \ValueVal)$\;
  }
}
    
\OutputSQL $\leftarrow$ \Transform(\AST, \Tables, \Columns, \Values) \tcp*{Generated AST created on-the-fly}
    \Return \OutputSQL\;
}
\end{algorithm}

Each set of \textit{Leaf nodes} can be represented as a quadruple for a given attribute (see line 7) which consists of the aggregator function position, table position, column position and value position. An example of such a quadruple is shown in Figure \ref{fig:data-flow-Working-pipeline2} (see the right side of the Root-node indicated as \texttt{Filter(2)}). We will focus on the leaf nodes surrounded by dotted-lines and green backgrounds. For instance, \texttt{Filter(2)} refers to the filter with position 2, which is equivalent to a query with an exact match filter. \texttt{A(0)} refers to an attribute without an aggregation function. \texttt{T(0)} and \texttt{C(1)} refer to the table with position 0, i.e. \emph{neighbors},  column with position 7 without aggregation function, i.e. \emph{neighbor\_mode}. Finally, \texttt{V(3)} refers to a value with position 3, i.e. the value 2.

This quadruple needs to be unpacked to extract the information about tables, columns and values as described informally above. Formally, the extraction of the tables, columns and values using the enhanced schema is described in lines 8 to 20 of Algorithm \ref{algo:GenereteSQLQueriesAlgo}. If a position of a certain table, column or value is not found in the keys of the hash map, the respective sampling function will select a new value within the constraints of the enhanced schema, e.g., \texttt{sampleTable()} for table sampling (see line 9). Then the corresponding hash map will add the new position-value pair. At the end of the loop, all hash maps are filled with required position-value pairs for tables, columns and values. 

Finally, the AST is created on the fly and the corresponding SQL is returned (see lines 21 and 22 of Algorithm \ref{algo:GenereteSQLQueriesAlgo}).

\subsubsection{Phase 3: SQL-to-NL Translation Phase.} 
In the third phase of our pipeline shown in Figure \ref{fig:data-flow-Working-pipeline1}, we generate the natural language questions (NLQs) that correspond to the newly generated SQL queries. To achieve this, we use \emph{GPT-3}~\footnote{Note that we also experimented with DBPal~\cite{weir2020dbpal} as an alternative but we opted for a custom pipeline using GPT-3 since the generated natural language questions are more fluent. However, DBPal can easily be integrated in our pipeline to further extend ScienceBenchmark with additional training data.}~\cite{brown2020gpt3}.  We present details on the evaluation of LLMs for generating NL questions given a SQL statement in Section~\ref{sec:SQL_to_NL_Model_Selection}. 

We fine-tuned GPT-3\footnote{\color{black}We use a fine-tuned version of GPT-3 to generate NL questions because fine-tuning GPT-4 is not available yet https://platform.openai.com/docs/guides/fine-tuning/what-models-can-be-fine-tuned.}} on a subset of 468 samples of the Spider training data for four epochs. This alleviates the need for prompt engineering. Thus, the input to GPT-3 is a SQL query, and we let GPT-3 generate 8 natural language question-candidates to increase the linguistic diversity. Since there is no additional input required beyond the SQL query (e.g., database schema, extra information about the DB, DB contents), this approach easily transfers to any new database without any manual effort or need for extra data.

For the more domain-specific databases, we conduct fine-tuning on GPT-3 with the manually created seed NL/SQL pairs. Afterwards we apply the fine-tuned LLMs to translate SQL to NL.
As in the Spider dataset, to obtain a larger variety of questions and achieve higher linguistic diversity, we generate several candidate NLQs per query.
This approach also approximates the Spider dataset where each SQL query has multiple semantically equivalent natural language questions.

\subsubsection{Phase 4: Discriminative Phase.} The last phase of our data generation pipeline, as shown in Figure \ref{fig:data-flow-Working-pipeline1}, selects the one or two best NL questions from the set of candidates generated in the previous phase. 

Consider, for instance, the following two NL questions depicted in Figure \ref{fig:data-flow-Working-pipeline1}: \emph{"Find the center object which has nearest neighbor with neighbor mode 2"} and \emph{"Find the center id of nearest neighbor object with neighbor mode smaller than 2"}. The discriminative phase aims at deciding, which one better represents the SQL query \begin{footnotesize}\texttt{"SELECT T1.objid FROM neighbors AS T1 WHERE T1.neighbormode = 2"}\end{footnotesize}.

Inspired by the centroid-based text summarization method \cite{rossiello-etal-2017-centroid}, the best NLQs are those, whose word embeddings are closest to the centroid of all sample questions. To find these points, we select the candidates that are closest to the centroid, i.e. the geometric median of all embeddings. For this, we apply \textit{SentenceBERT}\cite{reimers2019sentbert}, to generate a set of sentence embeddings for all candidates: $x_i \in \mathbb{R}^m$ and $1 \leq i \leq n$. The best NLQs are computed by taking the geometric median and selecting the closest embedding.

Consider a set \(X \in \mathbb{R}^m\) which contains \textit{n} embeddings of generated NL questions, \({x_1, x_2, ..., x_n}\), where \(m\) denotes the dimension of embedding space. By the definition of geometric median, we can find the closest embedding \(y \in X\) with respect to the centroid vector in the space by \emph{solving the optimization problem} formalized as $f(y)$: 

\begin{equation}
\label{eq: geometricMedian}
{\small
\begin{aligned}
    f(y) = \argmax_{y \in \mathbb{R}^m}{\sum^{n}_{i=1}{CosSim(x_i, y)}}
\end{aligned}}
\end{equation}
That is, finding the candidate NLQ whose embedding has the highest cosine similarity to the centroid. We perform this process $k$ times on the set $X \setminus \{y\}$ until we have the top $k$ natural language candidates. We choose one or two best NLQs, i.e., $k \in \{1, 2\}$.

\subsection{ScienceBenchmark Statistics}

\begin{table*}[t]
\small
\resizebox{0.6\textwidth}{!}{
    \begin{tabular}{@{} l r r r r r @{}}
\toprule
     \textbf{Dataset} & \textbf{Easy} & \textbf{Medium} & \textbf{Hard} & \textbf{Extra Hard} & \textbf{Total}\\ 
\midrule
     CORDIS Seed            & 4 (4\%) & 15 (15\%) & 38 (38\%) & 43 (43\%)& 100 \\
     CORDIS Synth              & 726 (55.59\%) & 494 (37.83\%) & 66 (5.05\%) & 20 (1.53\%) & 1306  \\     
     CORDIS Dev           & 25 (25\%) & 35 (35\%) & 19 (19\%) & 21 (21\%) & 100 \\ 

\midrule
     SDSS Seed              & 20 (20\%) & 54 (54\%) & 2 (2\%) & 24 (24\%) & 100 \\
     SDSS Synth                & 326 (15.82\%) & 1396 (67.73\%) & 138 (6.7\%) & 201 (9.75\%) & 2061 \\     
     SDSS Dev             & 12 (12\%) & 28 (28\%) & 20 (20\%) & 40 (40\%) & 100\\

\midrule
     OncoMX Seed            & \color{black}{34} (34\%) & \color{black}{33} (33\%) & \color{black}{19} (19\%) & \color{black}{14} (14\%)& \color{black}100 \\
     OncoMX Synth             & 464 (43.57\%) & 601 (56.43\%) & 0 (0\%) & 0 (0\%) & 1065\\
     OncoMX Dev           & \color{black}{21} (21\%) & \color{black}{32} (32\%) & \color{black}{27} (27\%) & \color{black}{20} (20\%)& \color{black}100 \\

\midrule \midrule
     Spider Train           & 1944 (22.45\%) & 2831 (32.7\%) & 1758 (20.3\%) & 2126 (24.55\%) &  8659 \\
     Spider Dev             & 250 (24.22\%) & 440 (42.64\%) & 174 (16.86\%) & 168 (16.28\%) & 1032 \\
\bottomrule
\end{tabular}
}
\caption{
New benchmark dataset called \emph{ScienceBenchmark} which we constructed using the automatic training data generation pipeline shown in Figure \ref{fig:data-flow-Working-pipeline1}. The size and complexity of the queries in the 3 databases of ScienceBenchmark are according to the Spider~\cite{yu-etal-2018-spider} hardness classification scheme. 
The datasets Seed and Dev are manually generated by domain and SQL experts. The datasets Synth are automatically generated. In the bottom part we also include the equivalent statistics of the Spider dataset for comparison. 
}
\label{tab:complexity_overview}
\end{table*}

Table~\ref{tab:complexity_overview} gives an overview of our new benchmark dataset called \emph{ScienceBenchmark} which we constructed using the automatic data generation pipeline shown in Figure \ref{fig:data-flow-Working-pipeline1}.
Note that for each of the three domain-specific databases described in Section~\ref{sec:databases}, we present two manually created subsets (Seed and Dev) and one automatically generated subset (Synth). The manually created Seed and Dev queries were created by a team of 20 domain and SQL experts as described in Section \ref{sec:manuel_data_collection}, while the Synth queries we produced by our data generation pipeline described in Section \ref{sec:automatic-generation}. The Seed queries are used for the automatic data generation pipeline to generate synthetic data (Synth), while the Dev queries are used to evaluate NL-to-SQL systems.



{\color{black}Table~\ref{tab:complexity_overview} also shows the query difficulty (a metric defined by the creators of Spider~\cite{yu-etal-2018-spider}) distribution for each dataset. For the CORDIS and SDSS datasets, we note that the complexity of the queries is higher than the queries in the Spider Dev Set. For OncoMX, the complexity of the queries is closer to that of the distribution of the Spider dataset. This is due to the database featuring recursive traversals of complex hierarchies of anatomical entities which is outside of the scope of current NL-to-SQL systems.}
Note that the complexities of the queries generated by our pipeline are generally lower than the complexity of the manually created training data, or the complexity of the Spider dataset. The reason is that with more complex templates the generated queries tend to be semantically incorrect. 


\section{Evaluating the Quality of ScienceBenchmark}

In this section, we evaluate the \emph{quality} of our new benchmark dataset \emph{ScienceBenchmark}.
The main objective is to answer the following two research questions:
\begin{itemize}
    \item \emph{Research question 1: How well do current methods work for translating SQL to NL?}
    \item \emph{Research question 2: What is the quality of the automatically generated synthetic data, i.e. NL/SQL pairs?}
\end{itemize}

In order to answers these questions, we first present our evaluation of four different LLMs for translating SQL to NL. The best-performing LLM will then be used to generate the synthetic data. 
Afterwards, we evaluate the correctness of the synthetic data for each of the three databases of ScienceBenchmark by performing an expert evaluation.

\subsection{Evaluation of {\color{black}LLMs} for SQL-to-NL Translation}
\label{sec:SQL_to_NL_Model_Selection}

This section describes the experiments we performed in order to decide which LLM to incorporate into our automatic data generation pipeline. We evaluate the accuracy of each LLM in isolation. The best LLM is then used in Phase 3 "SQL-to-NL Translation" of our automatic data generation pipeline shown in Figure \ref{fig:data-flow-Working-pipeline1}.

We analyze the performance of four different LLMs, which are all based on large-scale transformer language models~\cite{vaswani2017attention}. We use these LLMs for \emph{translating the SQL queries} in the Spider Dev set \emph{to natural language}. 
We apply various automated metrics to these results as well as an expert evaluation. 

\paragraph{Large Language Models} We have chosen the following four LLMs for our SQL-to-NL translation: 
\begin{itemize}
\item GPT-2: A fine-tuned GPT-2-large model~\cite{radford2019gpt2} with an auto-regressive decoder-only large pre-trained language model, which is well suited for text generation.
\item GPT-3-zero: A zero-shot GPT-3 Davinci model~\cite{brown2020gpt3}, which is a larger version of the GPT-2 model pre-trained on even more data.
\item GPT-3: A fine-tuned GPT-3 Davinci model, which is GPT-3 fine-tuned on NL/SQL pairs.
\item T5: A fine-tuned T5-base model~\cite{raffel2019t5}, which is an encoder-decoder-based pre-trained language model developed for machine translation.
\end{itemize}

We fine-tuned a GPT-2-large language model on the Spider training data for 20 epochs. The GPT-3 model was fine-tuned on a subset of Spider for 4 epochs\footnote{We decided to use only a subset of Spider to keep the costs low, since fine-tuning on all Spider data for 20 epochs would cost 600\$ (we only payed 10\$). Note that 4 epochs is the default value provided by GPT-3 for fine-tuning.}. For this, we sampled three NL/SQL-pairs from each database in the Spider training set at random, which resulted in a training set of 468 NL/SQL-pairs. We used a simple prompt to trigger the translation from SQL to NL. The T5-base model was fine-tuned on the entire Spider dataset, for 10 epochs. %

\begin{table}[t!]
\centering
\small
\resizebox{0.45\textwidth}{!}{
    \begin{tabular}{@{} l r r r r @{}}
\toprule
     \textbf{Metric} & \textbf{GPT-2} & \textbf{GPT-3-zero} & \textbf{GPT-3} & \textbf{T5}\\
\midrule
     \textbf{SacreBLEU}    & 33.85 & 30.36 & \textbf{38.55} & 31.79 \\
     \textbf{SentenceBERT} & 0.840  & 0.870 & \textbf{0.888} & 0.864 \\
     \textbf{Human Expert}  & 0.629 & \textbf{0.765} & 0.731  & 0.645 \\ 
\bottomrule
\end{tabular}
}
\caption{Evaluation of various {\color{black}LLMs} for generating natural language questions given a SQL query. The goal is to validate Phase 3 "SQL-to-NL Translation" of our automatic data generation pipeline shown in Figure \ref{fig:data-flow-Working-pipeline1}. The evaluation is performed on the Spider Dev set using two different automatic performance metrics (SacreBLEU and SententeceBERT) as well as human experts.}
\label{tab:sql2text_eval}
\vspace*{-15pt}
\end{table}

\paragraph{Metrics} Each LLM is evaluated using the SacreBLEU score~\cite{papineni2002bleu,post2018sacrebleu} and the SentenceBERT score~\cite{reimers2019sentbert} automatic metrics. SacreBLEU is an instantiation of the BLEU score which measures the word overlap between two sentences. However,  word overlap metrics do not capture semantically equivalent natural language questions. 

For instance, consider the following two sentences (1) "\emph{Find  all Starburst galaxies?}" and (2) "\emph{Return all the spectroscopically observed galaxies that lie in the starburst class.}". Both statements describe the same information request, however, they have a low BLEU score.  Thus, we also use SentenceBERT, which measures the semantic similarity of sentences. 

Additionally, since automated evaluations are not perfectly reliable, we also ran an expert evaluation where 7 SQL experts rated the generated questions. For each expert, we randomly sampled 25 SQL queries from the Spider Dev set and let each of the four LLMs generate the corresponding natural language question. Thus, each expert annotated 100 SQL/synthetic question-pairs. In other words, for each LLM, we have 175 expert annotations. 

\subsubsection{Results for Spider Datasets}

The evaluation results on the Spider Dev Set using various metrics are summarized in Table~\ref{tab:sql2text_eval}. 
The first two lines show the scores given by the automatic metrics SacreBLEU and SentenceBERT. The third line shows the evaluation by human experts. This metric shows the ratio of samples that human experts regarded as being correct. 

We observe that the GPT-3 model outperforms the other models by a large margin in terms of SacreBLEU score. The average SentenceBERT similarity is also highest for the fine-tuned version of GPT-3. The human expert evaluation shows that both versions of GPT-3 achieve significantly higher scores than GPT-2 and T5. However, the difference between the two versions of GPT-3 are not significant, i.e. 76.5\% vs. 73.1\%. Thus, we opt to use the fine-tuned version of GPT-3 since it achieved the highest scores on ScareBLEU and SentenceBERT.

\subsubsection{Results for ScienceBenchmark}

We also ran expert evaluations for each of the three domains contained in the \emph{ScienceBenchmark}. For this, we translated 100 manually generated SQL queries (called dev queries) to NL questions using a GPT-3 model, which was fine-tuned on the specific database. For each database, we used the manually created training queries and the same 468 Spider queries used above to fine-tune GPT-3. We then performed the expert evaluation for the domain-specific GPT-3 models. 

For the CORDIS dataset, GPT-3 correctly translates SQL to NL in $82\%$ of cases, for OncoMX $73\%$. For SDSS, the ratio is lower at $53\%$, which is mostly due to the higher complexity of the dev queries. 

\vspace{3pt}

\textbf{Answer to Research Question 1:} We have shown that LLMs are powerful enough to generate good NL questions for a variety of domains, be it common knowledge or highly domain-specific.

\subsection{Evaluation of Synthetic Datasets (Silver Standard) of ScienceBenchmark}

We now analyze the quality of the synthetic datasets (or silver standard) for the novel domains of our \emph{ScienceBenchmark} via an expert evaluation. Note that in the previous section we only evaluated the translation of SQL to NL for the \emph{manually written} Dev Set SQL queries. Now we evaluate the synthetic datasets of CORDIS, SDSS and OncoMX, where \emph{both the SQL queries and the corresponding NL questions are automatically generated} using the pipeline in Figure \ref{fig:data-flow-Working-pipeline1}.

Distantly labelled data, also known as "silver standard" data has been used as a resource for reliably training neural networks when manually labelled data or "gold standard" data is scarce or unavailable. As shown in previous work on distant supervision \cite{roller-stevenson-2015-making}, training data does not have to be perfect and neural networks can learn from noisy or partially incorrect training data.  

Many training data generation systems such as DBPal \cite{weir2020dbpal} are based on the principal that silver standard data (possibly noisy data), is sufficient for training. Although DBPal provides an end-to-end systems analysis to show the effectiveness of the generated data, they do not provide any manual analysis of the quality or accuracy of the training data itself. 

Because the SQL queries in our data generation pipeline are generated using rule-based algorithms and filtered with heuristics crafted by domain experts to ensure the domain relevance of the SQL queries, we evaluate the \emph{semantic equivalence} of the NL questions generated in our pipeline i.e. we check if the NL question matches the meaning of the SQL query. First, we randomly sampled 100 NL/SQL-pairs from each synthetic dataset (CORDIS, SDSS, OncoMx) proportionally in line with the Spider hardness classification schema. Afterwards, we manually evaluated the NL questions against the matching SQL query. 

{\color{black}Table \ref{tab:silver_standard_eval} shows the results of our manual evaluation of the synthetic NL/SQL pairs. First, we analyzed the \emph{semantic meaningfulness} of the generated SQL queries (see second column of Table \ref{tab:silver_standard_eval}). For instance, a query that applies an aggregation over numeric data values is considered meaningful, while applying an aggregation over string values is not. Our results demonstrate that we generated meaningful queries in 91 to 97\% of the cases in all three datasets. 

We also analyzed the \emph{semantic equivalence} of the generated NL question and the respective generated SQL query. The results show that in 75 to 83\% of the cases we observe a semantic equivalence.

In summary, the synthetic queries are automatically generated and can be considered \emph{silver standard data}. Previous experiments have shown that (even noisy) silver standard data outperform curated datasets (see \cite{honovich2022unnatural}). These  silver standard data can even be false. Hence, we apply the same approach and do not filter out any query.
}


\begin{table}[t!]
\centering
\begin{tabular}[h]{@{} l l l @{}}
\toprule
\textbf{Dataset} & \textbf{{\color{black}Semantic }}  & 
\textbf{Semantic Equivalence} \\ 
\textbf{}    & 
{\color{black}\textbf{Meaningfulness of SQL}}          &  \textbf{of NL and SQL} \\ 
\midrule
     CORDIS & {\color{black}97\%} & 83\% \\
     SDSS & {\color{black}97\%} & 76\%  \\
     OncoMX & {\color{black}91\%} & 75\% \\ 
\bottomrule
\end{tabular}
\caption{{\color{black}Manual evaluation of 100 randomly chosen synthetic NL/SQL-pairs} of ScienceBenchmark. The results show both {\color{black} the semantic equivalence of the automatically generated NL questions with their corresponding, automatically generated SQL queries and the semantic meaningfulness of these automatically generated SQL queries}.}
\label{tab:silver_standard_eval}
\vspace*{-15pt}
\end{table}

\vspace{2pt}

\textbf{Answer to Research Question 2:} {\color{black}This analysis demonstrates that the quality of the synthetically generated or "silver standard" data for the three novel domains of ScienceBenchmark is high. Moreover, as we will show in Section \ref{sec:experimental_results}, using the synthetic datasets for training also significantly improves the performance of NL-to-SQL systems evaluated on ScienceBenchmark.}

\section{Baseline Experiments: Using ScienceBenchmark to Evaluate NL-to-SQL Systems}


In this section, we perform baseline experiments to evaluate the performance of popular NL-to-SQL systems on \emph{ScienceBenchmark}. The main research questions we want to address are as follows:

\begin{itemize}
    \item \emph{Research question 3: How well do current NL-to-SQL systems perform on complex, real-world scientific databases?}
    \item \emph{Research question 4: How much can NL-to-SQL systems be improved using data augmentation when little to no domain-specific training data is available?}
\end{itemize}


\subsection{NL-to-SQL Systems}
To test the performance of NL-to-SQL systems on \emph{ScienceBenchmark}, we selected state-of-the-art fine-tuned and {\color{black} few-shot} systems. The fine-tuned systems meet the following criteria:
1) Access to the open source model and the pre-trained model weights;
2) Access to bidirectional conversion code between SQL and intermediate representations (IR) of the systems, if any IR is used in the model\footnote{Since the SQL-to-NatSQL \cite{gan-etal-2021-natural-sql} conversion code is not available, as announced by the author in their code repository,\url{https://github.com/ygan/NatSQL}, all systems integrated with NatSQL have been excluded from our experiments.}. 

Therefore, for our experiments with fine-tuned models we use three different state-of-the-art NL-to-SQL systems: the only two completely open source state-of-the-art NL-to-SQL systems from the Spider leaderboard\footnote{\url{https://yale-lily.github.io/spider}} (T5-Large, SmBoP) and an industrial-strength NL-to-SQL system, which has recently been extended to handle complex, real-world datasets (ValueNet).
\squishlist
\item
 T5-Large \cite{raffel2019t5} (with Picard \cite{scholak2021picard} for constrained decoding). T5-Large is a language model, which is pre-trained on large amounts of text data\footnote{Due to compilation issues with Picard's decoder architecture implemented in Haskell, we only used T5-Large without the Picard version of the code provided by Picard.}.
\item SmBoP \cite{rubin-berant-2021-smbop-semi} (with GraPPa \cite{yu2021grappa}). SmBoP implements a novel autoregressive bottom-up decoder, enhanced with the GraPPa, which is a language model specifically pre-trained for the NL-to-SQL task. 
\item ValueNet~\cite{brunner2021valuenet}. ValueNet is based on the IRNet~\cite{guo2019irnet} architecture and extends the SemQL grammar by adding values to the generated queries to make them executable. To handle the SDSS astrophysics data, we extended the SemQL grammar for ValueNet to incorporate mathematical operations~\footnote{We only incorporated the mathematical operations for ValueNet as it was straightforward to extend SemQL. The T5 architecture handles mathematical operations out-of-the-box as we use the unconstrained version.}.
\squishend


{\color{black}To evaluate the difficulty of ScienceBenchmark, we also include baseline experiments with two {\color{black}LLMs}, GPT-3.5 from OpenAI and for comparison a newly released open source model, LLaMA2 70B from Meta\footnote{https://ai.meta.com/llama/}.}

For reproducibility reasons, we provide both the source code of our automatic training data generation approach and the datasets of \emph{ScienceBenchmark}\footnote{The source code, the datasets, the hyperparameters, and the model-specific experiments' hardware specifications for ScienceBenchmark can be found at: \url{https://sciencebenchmark.cloudlab.zhaw.ch/}}. 

\subsection{Experimental Setup}
For each of the 3 databases of \emph{ScienceBenchmark}, we ran four experiments:
\squishlist
    \item \emph{Spider Train (Zero-Shot)}: Here, we train the NL-to-SQL systems on the \emph{Spider Train Set}, and run the evaluation on the \emph{Dev Set} of the respective new domain, i.e. on the evaluation set of CORDIS, SDSS and OncoMX. 
    \item \emph{Spider Train + Domain Train}: We train the NL-to-SQL systems on a mix of the Spider Train Set and the manually created training data, e.g. \emph{CORDIS Train}. The goal is to understand how much impact manually generated, domain-specific training queries have on the performance of the NL-to-SQL system. 
    \item \emph{Spider Train + Domain Synth}: We train the systems on a mix of Spider training data and the synthetic data, which we automatically generated using our training data generation pipeline. For instance, \emph{CORDIS Synth} is the automatically generated (synthetic) training data for the domain \emph{research policy making}.
    \item \emph{Spider Train + Domain Train + Synth}: Here, we train the systems on a mix of the Spider training data, manually created training data, and synthetic data from each domain. This shows the impact of using both synthetic and manually curated data. 
\squishend


{\color{black} The left-most column of Table \ref{tab:augment2} shows all possible settings, i.e., the four experiments per Dev Set (database), which is shown in the second column.} The evaluation is performed using \emph{Execution Accuracy}, which is the metric used by the Spider benchmark. In other words, we measure in how many cases the result sets of the predicted SQL queries correspond to the result sets of the Dev Set SQL queries. 



\subsection{Experimental Results}
\label{sec:experimental_results}

{\color{black} Table \ref{tab:augment2} is divided in two parts: The left most parts show the execution accuracy for NL-to-SQL systems with fine tuning using \emph{ScienceBenchmark} (for the experiments described in the previous section). The right most parts of the table show results for LLMs with prompt-engineering. } 

 {\color{black} Let us first discuss the results of the NL-to-SQL systems that we fine-tuned, i.e., ValueNet, T5-Large and SmBoP.}
Our results show that \emph{ScienceBenchmark} poses a challenge to current NL-to-SQL approaches. As expected, the performance of the various systems is very low in the zero-shot setting, as the domains covered in Spider are hardly transferable to the novel complex domains. {\color{black}The results also highlight that simply augmenting the number of training samples does not enable the models to achieve high accuracy (80\% and above) on the benchmark}. In fact, adding the manual and synthetic training data of our domains, increases the scores by a large margin, however, the absolute scores remain low. For instance, the score increases by 23\% on CORDIS when training ValueNet on all the data.  {\color{black} For T5-Large, data augmentation improves the zero-shot case by 45\% resulting in a maximum execution accuracy of 56\%. However, with an execution accuracy of 56\% the task is far from solved. This trend is noticeable for each of the three novel domains. We also observe that the particularities of each domain create a different degree of difficulty to each system as witnessed by the different scores each system achieves in each domain with zero-shot learning.

\begin{table*}[t!]
\centering
\small
\resizebox{0.95\textwidth}{!}{
    \begin{tabular}{@{} l l l l l || l l l @{}}
     \toprule
     \multicolumn{2}{@{}l}{\textbf{Dataset}} & \multicolumn{3}{l}{\color{black}{\textbf{Fine-Tuning with NL-to-SQL Systems}}} 
     & \multicolumn{2}{l@{}}{\color{black}{\textbf{Zero \& Few-shot-Prompting with LLMs}}} \\
     \cmidrule(r){1-2}
     \cmidrule(l){3-5}
     \cmidrule(l){6-8}
     \textbf{Train Set} & \textbf{Dev Set} & \textbf{ValueNet} & \textbf{T5-Large w/o Picard} & \textbf{SmBoP+GraPPa} & \color{black}{\textbf{Prompting*}} & \color{black}{\textbf{GPT-3.5 (175B)}} & \color{black}{\textbf{LLaMA2 (70B)}} \\ 
     \midrule
     Spider Train (Zero-Shot)   & CORDIS                   & 0.12          & 0.16          & 0.16  & \color{black}{Zero-shot} & \color{black}{0.57} & \color{black}{0.06}\\
     Spider Train + Seed CORDIS & CORDIS          & 0.20 (+0.08) & 0.20 (+0.04)   & 0.20 (+0.04) & \color{black}{One-shot} & \color{black}{0.60} & \color{black}{0.22}\\
     Spider Train + Synth  CORDIS & CORDIS        & 0.31 (+0.19) & 0.30 (+0.14)   & 0.23 (+0.07) & \color{black}{-} & \color{black}{-} & \color{black}{-}\\
     Spider Train + Seed CORDIS + Synth  CORDIS  & CORDIS & 0.35 (\textbf{+0.23}) & 0.29 (+0.13)   & 0.21 (+0.05) & \color{black}{Five-shot} & \color{black}{0.57} & \color{black}{0.18}\\
     \midrule
     Spider Train (Zero-Shot) & SDSS                        & 0.08          & 0.05          &  0.06 & \color{black}{Zero-shot} & \color{black}{0.29} & \color{black}{0} \\
     Spider Train + Seed SDSS & SDSS              & 0.11 (+0.03)  & 0.06 (+0.01)  &  0.10 (+0.04) & \color{black}{One-shot} & \color{black}{0.32} & \color{black}{0.06}\\
     Spider Train + Synth  SDSS  & SDSS           & 0.18 (+0.10)  & 0.12 (+0.07)  &  0.13 (+0.07) & \color{black}{-} & \color{black}{-} & \color{black}{-} \\ 
     Spider Train + Seed SDSS + Synth SDSS  & SDSS    & 0.21 (\textbf{+0.13})  & 0.15 (+0.10)  &  0.15 (+0.09) & \color{black}{Five-shot} & \color{black}{0.33} & \color{black}{0.03}\\ 
     \midrule
     Spider Train (Zero-Shot) & OncoMx                      & \color{black}{0.13}          & \color{black}{0.11}         & \color{black}{0.19} & \color{black}{Zero-shot} & \color{black}{0.50} & \color{black}{0.11}\\
     Spider Train + Seed OncoMx & OncoMx          &  \color{black}{0.44 (+0.31)} & \color{black}{0.35 (+0.24)} & \color{black}{0.32 (+0.13)} & \color{black}{One-shot} & \color{black}{0.51} & \color{black}{0.21}\\
     Spider Train + Synth OncoMx  & OncoMx        & \color{black}{0.23 (+0.10)} & \color{black}{0.27 (+0.14)} & \color{black}{0.20 (+0.01)} & \color{black}{-} & \color{black}{-} & \color{black}{-}\\ 
     Spider Train + Seed OncoMx + Synth OncoMx & OncoMx    & \color{black}{0.49} (+0.36) & \color{black}{0.56 (\textbf{+0.45})} & \color{black}{0.35 (+0.16)} & \color{black}{Five-shot} & \color{black}{0.55} & \color{black}{0.32}\\
     \bottomrule
\end{tabular}
}
\caption{Putting \emph{ScienceBenchmark} into practice: Evaluation of NL-to-SQL systems without and with data augmentation {\color{black}(left most parts of the table)}. Execution Accuracy is evaluated  on the Dev Set of three novel datasets. The numbers in brackets refer to the relative improvements with respect to the zero-shot baseline. {\color{black} The right most parts of the table show results for LLMs with prompt-engineering. } }
\label{tab:augment2}
\end{table*}

Let us now analyze the performance of LLMs where we use zero-shot and few-shot prompting as shown in the right part of Table \ref{tab:augment2}. In general, we can observe that GPT-3.5 performs better than LLaMA2, since the former has significantly more parameters (175B vs. 70B). Moreover, we can see that GPT 3.5 performs best on CORDIS reaching a maximum execution accuracy of 60\%. However, for SDSS, the dataset with the most numerical values and mathematical operators, the highest execution accuracy reaches only 33\%. Again, the particularities of each domain play a significant role in determining the performance of each model.

In summary, these results demonstrate that ScienceBenchmark is a highly challenging dataset and thus confirms that translating NL-to-SQL is far from being solved. }

\vspace{2pt}

\textbf{Answer to Research Question 3}: The results show that the current state-of-the-art approaches, which work exceptionally well on Spider, do not perform well on real-world databases. Thus, we pose this dataset as a challenge, which requires more than an increase in training data size. It requires novel approaches that are able to handle all the new complexities introduced by these domains.

\vspace{2pt}

\textbf{Answer to Research Question 4:} The results also show that for each domain and NL-to-SQL system, the combination of seed and synthetic queries yields an \emph{improvement over the zero-shot baseline of up to {\color{black}45\%}}. The magnitude of the improvements varies depending on the NL-to-SQL system and domain.  


\subsection{Discussion of the Experiments}
Our experiments show that our automated data augmentation pipeline creates training data which is well suited for bootstrapping novel domains. The magnitude of the improvement depends on the NL-to-SQL systems themselves. However, for all highly domain-specific databases, the performance of the systems trained with synthetic data improved. This is useful when adapting an NL-to-SQL system trained on Spider data for a novel and more complex domain for a database with more tables, columns and rows. 

The results highlight the necessity to work on real-world applications. In the zero-shot setting all the state-of-the-art systems achieved poor performance. For instance, in the SDSS domain none of the systems achieved an accuracy of even 10\%. Even with the fine-tuning data, the performance of the systems are far from the 70\% achieved in the Spider setting. Thus, the results show the need of a more complex and real-world oriented benchmark. 

Furthermore, our results reveal that the usage of the synthetic data is useful to increase the performance of fine-tuned models. In most cases using the large set of synthetic data alone yields better results than using the small manual dataset for training. The mix of both the synthetic and the manual data yield the best results.
\textbf{Thus, \emph{ScienceBenchmark} is highly challenging as it requires to adapt most systems to domain-specific knowledge and to handle complex, real-world scientific databases -- which are in stark contrast to the relatively simple databases of the Spider benchmark.} 


\section{Related Work}
In this section, we review the related work regarding \textit{data augmentation} and \textit{NL-to-SQL benchmarks}.
\subsection{Data Augmentation}
Due to the need of deep learning models for high volumes of training examples, combined with sparsity of training data and the cost of manually creating it, a lot of research has been done in the area of data augmentation.

Previous work on data augmentation for NL-to-SQL systems mainly focuses on generating SQL queries that run over a single table rather than over multiple tables of a complete relational database. 
One such example is DBPal \cite{weir2020dbpal}, a template-based approach for generating NL/SQL-pairs, which uses manually-crafted templates of NL/SQL-pairs, which can be filled with the names of tables and columns in order to create training instances.
Additionally, the authors propose NL augmentations such as paraphrasing, random deletions and synonym substitutions.
However this approach might create "robotic" NLQs whose quality might be reduced by the proposed augmentations, since designing rules that can consistently work across all possible questions is notoriously hard.

Another approach \cite{guo2018questiongeneration} creates SQL queries by using simple SQL templates and sampling column names and values from a given table and then applies Recurrent Neural Networks (RNNs) to generate the equivalent NLQ. 
Some key differences to our work are that: (i) we can generate augmented data without completely relying on manually created NL/SQL-pairs or templates and that (ii) our NLQ augmentation step is much more robust and can generate completely new and realistic NL utterances.

Another approach \cite{pingchuan2021mtteql} uses Metamorphic Rules (MRs) to create equivalent alterations of NLQs and database schemas from given NL/SQL/DB-triplets.
Even though this work mainly focuses on providing a more robust evaluation framework for NL-to-SQL systems, it also proposes a methodology for data augmentation that takes advantage of the MRs used for the evaluation.
More specifically, the authors present a set of MRs that can be used to create an alteration of either the NLQ or the database schema, while keeping them semantically equivalent to the original.
However, the proposed NLQ transformations are relatively simple (i.e., synonym substitution and prefix insertion/deletion/substitution) compared to our approach for generating novel and fluent NLQs.
Additionally, compared to our work, this approach is not capable of augmenting the SQL part of the training examples and requires a hand-crafted set of NL/SQL-pairs in order to work for a new database.

The more recent work presented in \cite{wu2021dataaugmentation} is one of the few proposed architectures that can generate examples that cover multiple tables of a relational database.
This work generates SQL queries by creating templates using an abstract syntax tree grammar and filling them with attributes from the database.
The NLQs are then generated using a hierarchical, RNN-based neural model, that recursively generates explanations for all parts of the queries and then concatenates them.
Our work differs from the previous because we consider much more complex and robust SQL-to-NL models that can create NLQs with much higher variety and fluency, since they are generated by taking the entire SQL query into account.

Finally, our work differs from all previous work in the sense that instead of simply increasing the performance on a generic dataset like Spider \cite{yu2019spider}, we focus on adapting an NL-to-SQL system on new, unseen and complex databases, with little to no manual effort.

\subsection{NL-to-SQL Benchmarks}

Progress in NL-to-SQL systems was systematically impeded by the lack of a common, large-scale benchmark dataset. The introduction of  WikiSQL \cite{zhong2017seq2sql} and Spider \cite{yu-etal-2018-spider}, has drastically changed the landscape, allowing for the introduction of deep learning techniques to tackle the problem, as well as providing a common benchmark for comparing different approaches.
These two benchmark datasets remain the main point of reference for NL-to-SQL systems despite much criticism (e.g.,  WikiSQL has low complexity and multiple errors \cite{hwang2019sqlova} and Spider databases are not realistic  \cite{hazoom2021sede}).

NL-to-SQL benchmarks can be classified into: \emph{domain-specific} and \emph{cross-domain} datasets.
Domain-specific datasets focus on a single database from a specific domain, such as: movies and television series (IMDb \cite{sqlizer}), restaurant and shop reviews (Yelp \cite{sqlizer} and Restaurants \cite{tang2000automated,popescu2003theory}), academic research (Scholar \cite{DBLP:conf/acl/IyerKCKZ17} and Academic \cite{nalir}), financial data (Advising \cite{finegan2018advising} and FIBEN \cite{sen2020athena++}), medical data (MIMICSQL \cite{wang2020medical}), and questions and answers from Stack Exchange (SEDE \cite{hazoom2021wildtext-to-sql}). 

In contrast, cross-domain datasets contain multiple databases, taken from different domains.
Spider-DK \cite{gan2021spidersyn} and Spider-Syn \cite{gan2021spidersyn}, are extensions of Spider which explore system capabilities at cross-domain generalization and synonym robustness.
KaggleDBQA \cite{lee2021kaggledbqa} is another cross-domain dataset, although of much smaller size, that has been extracted from Kaggle and features databases taken from the Web. Another cross-domain dataset is OTTA~\cite{deriu-etal-2020-methodology}, which uses an inverse annotation procedure, whereby automatically generated queries, which are visually displayed are annotated with natural language questions by non-SQL experts.

\textbf{The main difference of our novel benchmark \emph{ScienceBenchmark}}, compared to all aforementioned datasets, is twofold: (i) it contains scientific domains that use domain-specific vocabulary, and (ii) it was developed by scientists and domain-experts over the course of a multi-year research project including partners from both academia and industry. Hence, the deep interactions between scientists and domain experts ensured high quality examples that reflect queries posed by actual users of these complex, scientific databases. 

\section{Conclusions}

In this work, we introduce the novel benchmark \emph{ScienceBenchmark} for evaluating NL-to-SQL systems {\color{black} as well as LLMs} against complex, real-world scientific databases. We also show the end-to-end pipeline 
 for automatically generating large synthetic training datasets for highly domain-specific databases, which are more complex both in terms of the subject matter and in terms of the number of tables, columns and rows than the databases used in the Spider benchmark. 

\textbf{Our experimental results show that \emph{ScienceBenchmark} poses a significant challenge to current NL-to-SQL approaches {\color{black} as well as LLMs}}. While these systems work exceptionally well on the Spider dataset which has relatively simple databases, they do not perform well on complex, real-world scientific databases. Thus, we argue that \textbf{ScienceBenchmark can serve as a new baseline benchmark for evaluating NL-to-SQL systems {\color{black} as well as LLMs} and thus sets the stage for novel research efforts to handle the complexities introduced by these real-world challenges}.
 

\begin{acks}
This project has received funding from the European Union’s Horizon 2020 research and innovation program under grant agreement No 863410. We also thank Jonathan F\"urst and Farhad Nooralahzadeh for their contributions in evaluating large language models.
\end{acks}


\bibliographystyle{ACM-Reference-Format}
\bibliography{custom}

\end{document}